\documentclass[12pt]{iopart}
\usepackage{epsfig}
\usepackage{graphicx}
\usepackage{epstopdf}
\usepackage{color}

\newcommand{\be}{\begin{equation}}
\newcommand{\ee}{\end{equation}}
\newcommand{\bea}{\begin{eqnarray}}
\newcommand{\eea}{\end{eqnarray}}
\newcommand{\vv}[1]{\mbox{\bf #1}}

\newcommand{\dvp}{\delta\vv{p}}
\newcommand{\dvq}{\delta\vv{q}}

\begin{document}
\title[Spectra of Harmonium in a magnetic field...]
{Spectra of Harmonium in a magnetic field using an
initial value representation of the semiclassical propagator}
\author{Frank Grossmann}
\address{Institut f\"ur Theoretische Physik, Technische Universit\"at
Dresden, D-01062 Dresden, Germany}
\author{Tobias Kramer}
\address{Institut f\"ur Theoretische Physik, Universit\"at Regensburg, D-93053 Regensburg, Germany}
\date{\today}

\begin{abstract}
For two Coulombically interacting electrons in a quantum dot with 
harmonic confinement and a constant magnetic field, 
we show that time-dependent semiclassical calculations using 
the Herman-Kluk initial value representation of the propagator lead
to eigenvalues of the same accuracy as WKB calculations
with Langer correction. The latter are restricted to integrable systems,
however, whereas the time-dependent initial value approach allows for 
applications to high-dimensional, possibly chaotic dynamics and is 
extendable to arbitrary shapes of the potential.
\end{abstract}  

\submitto{\JPA}
\ead{frank@physik.tu-dresden.de}
\pacs{03.65.Ge, 03.65.Sq, 31.15-p}

\maketitle

\section{Introduction}
The problem of two interacting electrons in atoms,
ions and molecules is of fundamental importance in quantum mechanics,
the understanding of the Helium atom being one of the prime successes 
of the ``new quantum mechanics'' \cite{Hogaasen10}. If the Coulombic 
interaction between electrons and nucleus is replaced by a harmonic confining
potential, the system is referred to as Harmonium, quantum 
dot Helium or Hooke's law atom. After early work without
additional magnetic field \cite{KS62}, this problem, with and without a
magnetic field, has again come in the focus of interest
in the past 20 years mainly for two different reasons:

Firstly it is of principal interest to find (exact) analytic solutions
for a problem with genuine Coulomb interaction. That such 
exact solutions do exist has been shown for two electrons in a harmonic 
confinement in 3 dimensions, without an additional magnetic field 
\cite{KS62,Tau93}, and in 2 dimensions with a magnetic field by 
Ver\c{c}in \cite{Vercin91} and Taut \cite{Tau94}. The existence of 
closed solutions for specific ratios of the Coulomb interaction and
the harmonic potential is related to an accidental (hidden) symmetry visible
after mapping the Coulomb interaction to a four dimensional harmonic 
oscillator potential \cite{Turbiner94}.
Furthermore, in 2 dimensions and using perturbation theory for the
Coulomb interaction, analytic solutions have been given in \cite{MHW91}.
The analytic solutions can be put to good use in the testing of 
density functionals and other methods for many-body systems 
\cite{CP00,LG10,Kramer2010c}. Secondly, due to recent 
progress in nanofabrication, few electron quantum dots have come into 
the limelight as they may provide a realization of a quantum bit 
\cite{Turton95}.
 
Due to the fact that exact analytic solutions of the Sch\"odinger
equation are limited to an infinite set of discrete oscillator 
frequencies \cite{Tau93,Tau94}, also approximate analytic solutions have been
sought for. Apart from the perturbation theoretic ones in 2d,
mentioned above, also semiclassical approaches have been taken which
rely on the Wentzel-Kramers-Brillouin (WKB) \cite{Rosas00}, respectively the 
Einstein-Brillouin-Keller (EBK) quantization rules \cite{Larkoski06}.
Most notably, in 2d and with an additional magnetic field
this has been done analytically by Klama and Mishchenko \cite{KM98}. It turned
out that the problem of two electrons in 2d, although one of
4 degrees of freedom (DOF), is highly separable and therefore in 
the end only a 1 DOF WKB approach is needed, which describes the
spectrum with a high accuracy.
In another  semiclassical analysis of the harmonium problem, including 
a magnetic field, dimensional effects have been discussed by Nazmitdinov 
and collaborators \cite{NSR02}. 

Whereas in WKB energy information is
directly available, using the semiclassical initial value representation 
(IVR) of the quantum mechanical 
propagator of Herman and Kluk (HK) 
\cite{He81,HK84,Kay941} spectra can be gained by
Fourier transformation of a time series, usually an auto-correlation
function \cite{He812}.  In \cite{HK07} it has been shown that accurate
spectra for collinear Helium can be obtained in this way.
Furthermore, semiclassical work on one electron systems
in external laser fields gives the explanation for the plateau formation 
in high harmonic spectra \cite{vdSR00}.

In this paper we will show that the time-dependent semiclassical
approach, complementary to the energy-domain WKB method, is also capable
of reproducing high quality spectra for 2d Harmonium in a
homogeneous magnetic field. Thus adding to the knowledge of semiclassics
for interacting many particle dynamics is not a mere exercise due to the fact
that the time-dependent approach is not principally restricted in
dimensionality nor to the case of a circular quantum dot
and also additional time-dependent potentials
may be treated in a similar fashion without much more effort.
In Sec.\ \ref{sec:ham}, the Hamiltonian is briefly introduced,
whereas in Sec.\ \ref{sec:scivr} we review the semiclassical
IVR of Herman and Kluk based on frozen Gaussian wavepackets. The new results 
we gained are compared with full quantum as well as with (Langer-corrected) 
WKB results in Sec.\ \ref{sec:spec} and we give conclusions and an outlook
in Sec.\ \ref{sec:conc}.
 
\section{The Hamiltonian}
\label{sec:ham}

The Hamiltonian for the problem of two charged particles (charge $q$) in a 
2d circular quantum dot (dielectric constant $\epsilon$) 
with a confinement frequency of $\omega_0$ inside a magnetic field, 
derived from a vector potential $\vv A(\vv r_i)$ is given by
\be
H=\sum_{i=1}^2\left(\frac{(\vv p_i-q\vv A(\vv r_i))^2}{2m^\ast}
+\frac{1}{2}m^\ast\omega_0^2\vv r_i^2\right)
+\frac{\kappa}{|\vv r_1-\vv r_2|}.
\ee
Here $\kappa=q^2/(4\pi\epsilon\epsilon_0)$ and for two 
electrons $m^\ast$ is the (effective) mass and the charge is
$q=-e$. We are not explicitly 
taking into account electron spin, thereby neglecting the Zeeman energy of the spins in 
the magnetic field, see also \cite{NSR02}. In addition, as will be seen and commented 
on below (see Section 4.2), for the propagation in time, we will consider an 
unsymmetrized position-dependent part of the quantum mechanical wavefunction.

To make progress, center of mass (cm) and relative coordinates are
introduced, according to
\be
\vv R=\frac{1}{2}(\vv r_1+\vv r_2)\qquad \vv P=(\vv p_1+\vv p_2),
\ee
respectively
\be
\vv r=\vv r_1-\vv r_2\qquad \vv p=\frac{1}{2}(\vv p_1-\vv p_2).
\ee
In their terms, the Hamiltonian is separable (i.\ e.\ depends
on the sum of two terms which depend only on cm or relative coordinates,
respectively) and for $\vv A(\vv r_i)$ depending linearly on
coordinate reads
\be
H=H_{\rm cm}+H_{\rm rel}
\ee
with
\bea
H_{\rm cm}&:=&\frac{1}{2M}\left(\vv P+2e\vv A(\vv R)\right)^2
+\frac{1}{2}M\omega_0^2\vv R^2
\\
H_{\rm rel}&:=&
\frac{1}{2\mu}\left(\vv p+\frac{e}{2}\vv A(\vv r)\right)^2
+\frac{1}{2}\mu\omega_0^2\vv r^2+\frac{\kappa}{r},
\eea
where $M=2m^\ast$ is the sum of the two masses and 
$\mu=m^\ast/2$ is the reduced mass.

A constant magnetic field in the $z$ direction is derivable from
a vector potential
\be
A_x=-yB/2,\qquad A_y=xB/2,\qquad A_z=0
\ee
given in Cartesian coordinates and in symmetric gauge.
For the relative motion Hamiltonian in 2d this leads to
\be
\label{eq:hrel}
H_{\rm rel}(\vv p,\vv r)=
\frac{1}{2\mu}\left(p_x^2+p_y^2\right)+
\frac{1}{2}\mu\Omega^2\vv r^2+\omega_L L_z+\frac{\kappa}{r},
\ee
where $L_z=p_y x-y p_x$ is the $z$-component of the angular momentum,
$\omega_L=eB/(2m^\ast)$ is the Larmor frequency and 
$\Omega^2=\omega_0^2+\omega_L^2$.

In the following we will exclusively concentrate on the
relative motion, due to the fact that the solution of the
cm problem is trivial (it can easily be gained from the solution of
the relative motion by neglecting the Coulomb term and a suitable
rescaling). 

\section{A semiclassical initial value propagator}
\label{sec:scivr}

Following early work by Heller \cite{He81}, who first presented a 
semiclassical integral expression for a time-evolved wavefunction
in terms of fixed width (frozen) Gaussian wavepackets,
Herman and Kluk have derived the semiclassically correct prefactor for an 
$N$-degree of freedom system \cite{HK84}. The resulting
expression for the propagator in the case $N=2$ is
\bea
\label{eq:HK1}
K^{\rm HK}(\vv r,t;\vv r',0)
&=&\int\frac{{\rm d}^2p' {\rm d}^2q'}{(2\pi\hbar)^2}
\sqrt{\det {\bf h}}\langle \vv r|g(\vv p_t,\vv q_t)\rangle
\nonumber
\\
&&\exp\left\{\frac{i}{\hbar}S(\vv p',\vv q',t)\right\}
\langle g(\vv p',\vv q')|\vv r'\rangle,
\eea
where 
\be
S(\vv p',\vv q',t)=\int_0^t {\rm d}t' [\vv p^{\rm T}\dot{\vv q}
-H_{\rm rel}(\vv p,\vv q)] 
\ee
is the classical action. With a series of papers in 1994 
\cite{Kay941,Kay942,Kay943}, Kay has laid the ground for a flurry of 
publications using the Herman-Kluk propagator.
One of its main features is the integration over
multiple Gaussian wave functions 
\be
\label{eq:froz}
\langle \vv r|g(\vv p,\vv q)\rangle=\left(\frac{\det{\bf \gamma}}{\pi^N}\right)^{1/4}
\exp\left\{-\frac{1}{2}(\vv r-\vv q)^{\rm T}{\bf \gamma}(\vv r-\vv q)+
\frac{i}{\hbar} \vv p^{\rm T}(\vv r-\vv q)\right\}
\ee
with constant width parameter matrix $\gamma$ (therefore the term
``frozen'' is frequently used), which we here assume to be the
same for the initial and the final Gaussian.  

Furthermore, the matrix 
${\bf h}$, whose determinant appears in the preexponential factor of the 
semiclassical propagator is given by
\be
\label{eq:pre}
{\bf h}(\vv p',\vv q',t)=
\frac{1}{2}({\bf m}_{11}+\gamma{\bf m}_{22}\gamma^{-1}
-i\hbar\gamma{\bf m}_{21}-\frac{1}{i\hbar}{\bf m}_{12}\gamma^{-1})
\ee
for diagonal $\gamma$-matrix, and contains sub-blocks of the 
so-called stability (or monodromy) matrix to be discussed in more detail
below. In a numerical implementation, the branch of the square root  
in Eq.\ (\ref{eq:HK1}) has to be chosen in such a fashion that the whole 
expression is a continuous function of time \cite{Kay941}.

Classical dynamics enters the HK 
IVR of the propagator via the classical trajectories
$(\vv p_t=\vv p(\vv p',\vv q',t),\vv q_t=\vv q(\vv p',\vv q',t))$, that
are initial value solutions of Hamilton's equations, and which in the case
of the relative motion of 2 electrons in a quantum dot read
\bea
\dot q_x&=&\frac{1}{\mu}p_x-\omega_L q_y
\\
\dot q_y&=&\frac{1}{\mu}p_y+\omega_L q_x
\\
\dot p_x&=&-\omega_Lp_y-\mu\Omega^2q_x+\frac{e^2q_x}{q^3}
\\
\dot p_y&=&\omega_Lp_x-\mu\Omega^2q_y+\frac{e^2q_y}{q^3}.
\eea
An in depth study of the classical dynamics of two interacting particles
in a magnetic field in two dimensions has been given by Curilef and Claro
\cite{CC97}. Even without the harmonic confining potential, and in spite
of the presence of the repulsive Coulomb interaction, the motion 
is bound due to the presence of the magnetic field.

Furthermore, also the equations of motion for the stability matrix 
are classical equations. They will be given
in vector form and for $N=2$ we define $\dvp'=(\delta p_x',\delta p_y'),
\dvq'=(\delta q_x',\delta q_y')$ as small initial 
deviations in phase space. Their time evolved counter parts are
\bea
\label{eq:smat1}
\dvp_t&=&{\bf m}_{11}\dvp'+{\bf m}_{12}\dvq'
\\
\label{eq:smat2}
\dvq_t&=&{\bf m}_{21}\dvp'+{\bf m}_{22}\dvq',
\eea
where the $2\times 2$ matrices ${\bf m}_{ij} (i,j=1,2)$ 
are submatrices of the stability (or monodromy) matrix
\be
{\bf M}\equiv
\left(\begin{array}{cc}
{\bf m}_{11}&{\bf m}_{12}\\
{\bf m}_{21}&{\bf m}_{22}
\end{array}\right)
\equiv
\left(\begin{array}{cc}
\frac{\partial\vv p_{t}}{\partial\vv p'^{\rm T}}&
\frac{\partial\vv p_{t}}{\partial\vv q'^{\rm T}}\\
\frac{\partial\vv q_{t}}{\partial\vv p'^{\rm T}}&
\frac{\partial\vv q_{t}}{\partial\vv q'^{\rm T}}
\end{array}\right).
\ee
The equation of motion for  ${\bf M}$ can be obtained 
by linearizing Hamilton's equations for the deviations and reads
(for more details, see, e.g., App. 2C in \cite{Gross})
\be
\label{eq:stab}
\frac{\rm d}{{\rm d}t}{\bf M}=
\left(
\begin{array}{cc}
-\frac{\partial^2 H_{\rm rel}}{\partial \vv q_t \partial \vv p_t^{\rm T}}&
-\frac{\partial^2 H_{\rm rel}}{\partial \vv q_t \partial \vv q_t^{\rm T}}
\\
\frac{\partial^2 H_{\rm rel}}{\partial \vv p_t \partial \vv p_t^{\rm T}}&
\frac{\partial^2 H_{\rm rel}}{\partial \vv p_t \partial \vv q_t^{\rm T}}
\end{array}
\right)
{\bf M}=-{\bf J}{\bf H}{\bf M}.
\ee
Here the skew symmetric matrix
\be
{\bf J}=
\left(
\begin{array}{cc}
{\bf 0}&{\bf 1}
\\
-{\bf 1}&{\bf 0}
\end{array}
\right)
\ee
and the Hessian,  i.\ e.\ the matrix containing the second 
derivatives of the Hamiltonian $H_{\rm rel}(\vv p_t,\vv q_t)$,
\be
{\bf H}\equiv
\left(
\begin{array}{cc}
\frac{\partial^2 H_{\rm rel}}{\partial \vv p_t \partial \vv p_t^{\rm T}}
&
\frac{\partial^2 H_{\rm rel}}{\partial \vv p_t \partial \vv q_t^{\rm T}}
\\
\frac{\partial^2 H_{\rm rel}}{\partial \vv q_t \partial \vv p_t^{\rm T}}
&
\frac{\partial^2 H_{\rm rel}}{\partial \vv q_t \partial \vv q_t^{\rm T}}
\end{array}
\right)
\ee
have been used. We note that the Hessian of the Hamiltonian
(\ref{eq:hrel}), in contrast to the standard case 
of $H=T(\vv p)+V(\vv q)$, for  $\omega_L\neq 0$ 
also contains nonzero mixed second derivatives.

The initial conditions follow from the definition of the stability 
matrix to be
\be
{\bf M}(0)=
\left(
\begin{array}{cc}
{\bf m}_{11}(0)&{\bf m}_{12}(0)
\\
{\bf m}_{21}(0)&{\bf m}_{22}(0)
\end{array}
\right)=
\left(
\begin{array}{cc}
{\bf 1}&{\bf 0}
\\
{\bf 0}&{\bf 1}
\end{array}
\right)
\ee 
and in the numerics we solve the stability equations along with the
trajectories by using a symplectic integrator of second order, the
so-called symplectic leap frog (or position Verlet) algorithm \cite{GNS94}.
This automatically ensures that the action is discretized using a 
mid-point rule, which in general is necessary in the presence of a 
vector potential \cite{Schu}.

The HK propagator is determined entirely with the solution 
of classical initial value problems, hence it is frequently referred to 
as an initial value representation, in contrast to the well-known van 
Vleck-Gutzwiller propagator \cite{VV28,Gu67} that is based on the solutions 
of a classical root search problem (double sided boundary value problem).
In most numerical applications published so far, 
the bare propagator has been used to evolve a Gaussian wavefunction in
time (see also below). Applying the propagator 
to such a wavefunction, centered around $(\vv p_\alpha,\vv q_\alpha)$ and
with the same width parameter as the frozen Gaussians (\ref{eq:froz})
used for the propagator, according to
\be
\label{eq:GW}
\Psi_{\alpha}(\vv r,t)
=\int {\rm d}^3r' K^{\rm HK}(\vv r,t;\vv r',0)
\langle \vv r'|\Psi_\alpha(0)\rangle
\ee 
is eased by the fact that the integral over $\vv r'$ can be done analytically
by using
\bea
\label{eq:over}
\langle g(\vv p',\vv q')|\Psi_\alpha(0)\rangle&=&
\exp\Bigl\{-\frac{1}{4}(\vv q'-\vv q_\alpha)^{\rm T}\gamma(\vv q'-\vv q_\alpha)
\nonumber
\\
& &
+\frac{i}{2\hbar}(\vv q'-\vv q_\alpha)^{\rm T}(\vv p'+\vv p_\alpha)
\nonumber
\\
& &
-\frac{1}{4\hbar^2}(\vv p'-\vv p_\alpha)^{\rm T}\gamma^{-1}(\vv p'-\vv p_\alpha)\Bigr\}
\eea
for the overlap between the initial Gaussian wavefunction and the 
frozen Gaussian. After inserting the propagator (\ref{eq:HK1}) 
together with Eq. (\ref{eq:over}) into Eq. (\ref{eq:GW}), 
an integration over initial phase space (the initial conditions of the
classical trajectories) remains to be done. 
The sampling of the initial phase space is done using the Monte-Carlo
integration technique \cite{KHD86} and facilitated by the exponential 
damping of the integrand far away from the center of the initial wavefunction.
In the numerical results to be presented below the convergence of the
time signals was checked by increasing the number of propagated trajectories.
For four phase space degrees of freedom typically 10$^6$ trajectories
are needed for converged results. In cases of strongly chaotic dynamics
(not considered here) and if long-time information is needed this number 
may, however, increase dramatically \cite{pre09}.
We note in passing that in contrast to the
multi-trajectory Herman-Kluk method \cite{jpa02}, the single trajectory
Thawed Gaussian Wavepacket method \cite{He75} uses just a
single trajectory to express the final wavefunction.

\section{Semiclassical spectra}
\label{sec:spec}

In the following we compare spectra that have been gained by
two different methods, the energy-domain WKB method and the
time-domain IVR method with subsequent Fourier transform, to
full quantum mechanical ones.

\subsection{WKB spectra}
In the case of a circular quantum dot in two dimensions that we
consider, the semiclassical WKB approach becomes particularly attractive, 
due to the fact that also the relative motion is separable in 
polar coordinates \cite{KM98}. 

In the simple case without the Coulomb term,
a one DOF WKB quantization of the radial motion leads to the 
Fock-Darwin energies \cite{Fock28,Dar31,KM98,NSR02}
\be
E_{\rm rel}(n_r,m)=(2n_r+|m|+1)\hbar\Omega-m\hbar\omega_L
\ee
where
\be
n_r=0,1,2,\dots\qquad m=0,\pm 1,\pm 2,\dots
\ee
are the radial, respectively azimuthal quantum numbers.
We note in passing that these eigenvalues have originally been gained for a 
{\it single} electron in a magnetic field and an
additional harmonic confinement. In the case 
$\omega_0=0$, i.\ e.\ without the external confinement, 
for $m$ positive or zero, the spectrum 
reduces to the so-called Landau levels $E_{\rm rel}=(2n_r+1)\hbar\omega_L$,
which are infinitely degenerate with respect to the angular
momentum quantum number, see also \cite{Rosas00}.

Separability is still given, also with the Coulomb term. The 
WKB energies can then be gained by solving for the
roots of a quartic equation and numerically integrating a
complete elliptic integral \cite{KM98,NSR02}. 
We stress that in the course of the derivation the two-dimensional 
Langer modification \cite{La37} of the azimuthal (magnetic) quantum number 
$m^2-1/4\to m^2$ has to be performed.

Graphical
representations of the eigenvalues as a function of magnetic
field strength can, e.\ g., be found in \cite{MHW91,KM98,NSR02}.
It is worthwhile to note that as the magnetic field increases, 
the ground state shifts to levels with higher angular momentum
because the (repulsive) Coulomb energy gets smaller when the
angular momentum grows along with the average distance
between the electrons \cite{KM98}. Here, we give a collection of 
some eigenvalues in table \ref{tab:wkb}, where by looking at the results with
the Coulomb potential, one can see that for the magnetic field 
chosen, the $m=1$ state is now the ground state. 
For the numerics in this paper, for reasons of simplicity, we 
made all quantities dimensionless by setting
$\mu=\hbar=e=\kappa=1$ (please note that this does not correspond to
atomic units).
\begin{table}
\begin{tabular}{c|c|c|c|c|c|c}
&\multicolumn{3}{c|}{without Coulomb}&\multicolumn{3}{c}{with Coulomb}\\
$n_r$&$m=0$&$m=1$&$m=2$&$m=0$&$m=1$&$m=2$
\\
\hline
0&1.41&1.83&2.24&3.14&2.84&3.02
\\
1&4.24&4.66&5.07&5.75&5.55&5.78
\end{tabular}
\caption{(Dimensionless) Fock-Darwin WKB energies ($\kappa=0$) and 
WKB energies with the Coulomb repulsion ($\kappa=1$) for 
$\omega_L=\omega_0=1$.}
\label{tab:wkb}
\end{table}

\subsection{Spectra from time-series}

Now we turn to the determination of energies from time series.
By using the semiclassical propagator an auto-correlation function
can be calculated, according to
\be
c(t)=\langle \Psi(0)|\Psi(t)\rangle.
\ee
This time series can then be Fourier-transformed into the energy domain.
Using the expansion of the wavefunctions in (orthogonalized) energy eigenstates
$|\Psi(t)\rangle=\sum_nc_n|n\rangle\exp\{-iE_nt/\hbar\}$, we get
\bea
S(\omega)&\stackrel{!}{=}&\frac{1}{2\pi\hbar} 
\int {\rm d}t e^{i\omega t}{c}(t)
\\
&=&\sum_{n=0}^\infty|c_n|^2\delta(E_n-\hbar \omega).
\eea
The peaks of the spectrum $S(\omega)$ are thus located at the eigenvalues
of the Hamiltonian \cite{He812,Gross}. 
In principal also a nonlinear procedure called harmonic inversion can 
be used to extract the spectral information. This has the advantage 
that shorter time signals and therefore less trajectories in the 
semiclassical calculations are needed \cite{cpl97}.

\begin{figure}[htb]
\begin{center}
\includegraphics[width=0.75\linewidth]{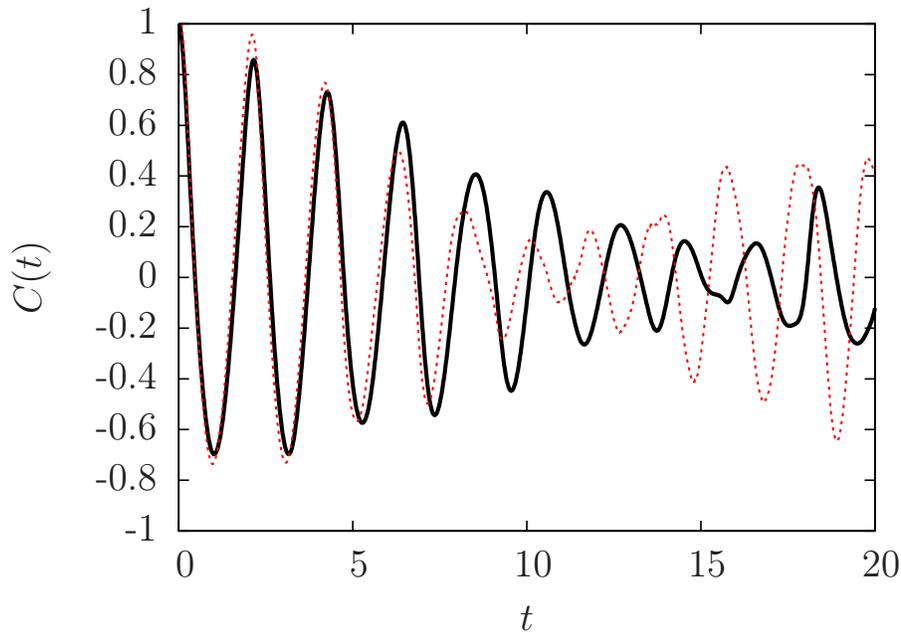}
\caption{Part of the auto-correlation time-series used to
extract spectral information for 2d harmonium in a magnetic
field ($\omega_L=\omega_0=1$) and with Coulomb interaction.
Solid line: full quantum, dotted line IVR semiclassical result.}
\label{fig:ts}
\end{center}
\end{figure}

We have performed the procedure just outlined both semiclassically with the
HK propagator as well as fully quantum
mechanically, using a finite difference method \cite{GCA84} and 
Cartesian coordinates for the two DOF.
A comparison of the two different time series for an
initial Gaussian wavepacket
\be
\Psi(\vv r,0)=\left(\frac{2\alpha}{\pi}\right)^{1/2}
\exp\{-\alpha(\vv r-\vv q_\alpha)^2+i\vv p^{\rm T}(\vv r-\vv q_\alpha)\}
\ee
with $q_{\alpha,x}=1,q_{\alpha,y}=0$, 
$p_{\alpha,x}=0,p_{\alpha,y}=-1$ and $\alpha=0.25$
is given in Fig.\ \ref{fig:ts}. Please note that we are not using 
an (anti-)symmetrized state here. Then both singlet and 
triplet states can be extracted from a single propagation without 
taking electron spin into account explicitly, see also \cite{SC05}.

We stress that without the Coulomb term the IVR results would be
numerically exact and on top of the quantum ones (not shown).
In the IVR case, with $\kappa=0$, even a single trajectory 
calculation according to the Thawed Gaussian approximation \cite{He75}
is sufficient to generate the time-dependent semiclassical results.
Therefore, the interesting case is the one with the Coulomb term.

For $\kappa=1$, the results for some eigenvalues extracted from the
time-series shown above are listed in Table \ref{tab:vgl}
and are compared to the corresponding WKB results.
\begin{table}
\begin{tabular}{c|c|c|c}
$m$&WKB&IVR&QM
\\
\hline
0&3.14&3.14&3.03
\\
1&2.84&2.84&2.83
\\
2&3.02&3.02&3.03
\end{tabular}
\caption{Comparison of WKB energies for $n_r=0$ and different values 
of $m$ with those from semiclassical
IVR and full quantum calculations, including the Coulomb interaction
and for $\omega_L=\omega_0=1$.}
\label{tab:vgl}
\end{table}
We note that the IVR and WKB semiclassical results are coinciding
to within the given accuracy, although in the time-dependent
calculations we did not(!) employ any Langer correction of the potential.
For single electron dynamics in the Coulomb potential the Langer correction,
necessary in the energy domain \cite{YU30}, is
needed in the time-domain if one uses non-Cartesian coordinates
\cite{ME94} and seems to be unnecessary for Cartesian coordinates
\cite{SBetal94,vdSR991}, which is corroborated here. Furthermore, with  better than
1 percent accuracy both semiclassical results are lying on top of the full 
quantum result, except for the $m=0$ eigenvalue. In the specific case considered,
the full quantum result is around 3 percent different from the
semiclassical ones for $m=0$. Furthermore, by coincidence, for the
present parameters, two quantum eigenvalues for $m=0$ and $m=2$
are degenerate. They do split up for a different
choice of the ratio $\omega_L/\omega_0$, however, where
a similar agreement between the differently calculated energy eigenvalues
is found (not shown).
The marked difference (a few percent) between the full quantum result 
for $m=0$ and the semiclassical ones does persist. 

\section{Conclusions and Outlook}
\label{sec:conc}

We have shown that the time-dependent semiclassical IVR methodology of
Herman and Kluk, based on Cartesian coordinates, is able to capture 
{\it spectral features} of two-electron quantum dots to a similar degree 
of accuracy as the WKB approach without the need to employ a Langer correction, 
complementing
work on the comparison of WKB {\it tunneling probabilities} with Fourier
transformed time-dependent semiclassical results \cite{KM94,cpl95}. The semiclassical
IVR method does not suffer from the restriction of WKB to integrable
(e.\ g.\ one DOF) systems, however. This allows the investigation of 
non-circular quantum  dots and also quantum dots in 3d or with more than 
two electrons.
For the study of the dynamics of nuclear degrees of freedom the HK method,
in the meantime, has become a widely used tool in the physics and 
chemistry communities (for an early review, see, e.\ g.\ \cite{camp99}).
An open question in the application to more than two-electron systems is the
problem of anti-symmetrization of the wavefunction, however.
Furthermore, we stress that there is no restriction on the value of 
$\Omega$, as is 
the case for the exact analytical solutions of the problem presented 
by Taut \cite{Tau94}, where the sequence of admissible $\Omega$ starts 
with a value on the order of one and converges to zero.

\ack The authors like to thank Eric J. Heller, Jan-Michael Rost
Lawrence S. Schulman, Dmitrii V. Shalashilin, and Steven Tomsovic, 
for fruitful discussions. 
FG thanks the Deutsche Forschungsgemeinschaft
for financial support through grant GR1210/4-2 and TK was supported by
the Emmy-Noether program KR 2889/2. This work was 
partially supported by the US National Science Foundation 
through a grant for the Institute for Theoretical Atomic,
Molecular and Optical Physics at Harvard University and Smithsonian
Astrophysical Observatory.

\section*{References}

\bibliographystyle{prsty}

\end{document}